\documentclass[useAMS,usenatbib,usegraphicx]{mn2e}
\usepackage{psfig}   
\usepackage{graphicx}
\usepackage{subfigure}

\title[Do O-stars form in isolation?]{Do O-stars form in isolation?}

\author[R.~J.~Parker and S.~P.~Goodwin]{
  Richard J.~Parker$^1$\thanks{r.parker@sheffield.ac.uk}
  and Simon P. Goodwin$^1$\thanks{s.goodwin@sheffield.ac.uk} \\
   $^1$ Department of Physics and Astronomy, University of Sheffield,
    Sheffield, S3 7RH, UK }

\begin{document}

\date{}
                             
\pagerange{\pageref{firstpage}--\pageref{lastpage}} \pubyear{2007}

\maketitle

\label{firstpage}

\begin{abstract}

Around 4\% of O-stars are observed in apparent isolation, with no associated
cluster, and no indication of having been ejected from a nearby cluster.  We
define an isolated O-star as a star $> 17.5 M_\odot$ in a cluster with total
mass $<100 M_\odot$ which contains no other massive ($>10 M_\odot$) stars.
We show that the fraction of apparently isolated O-stars is reproduced when
stars are sampled (randomly) from a standard initial mass function and a
standard cluster mass function of the form $N(M) \propto M^{-2}$.

This result is difficult to reconcile with the idea that there is a
fundamental relationship between the mass of a cluster and the mass of the
most massive star in that cluster.  We suggest that such a relationship is a
typical result of star formation in clusters, and that `isolated O-stars'
are low-mass clusters in which massive stars have been able to form.

\end{abstract}

\begin{keywords}   
stars: formation -- stars:mass function -- open clusters and associations
\end{keywords}

\section{Introduction}

It is believed that the vast majority of stars in the Galaxy form
in clusters with masses of a few $10$s to $\sim 10^5
M_\odot$ \citep{Lada03}.  The stars in these clusters
appear to form with an almost universal initial mass function 
\citep[see][]{Kroupa02, Kroupa07, Chabrier03}.  

For a typical initial mass function it is thought that one massive O-star 
forms per $200$ -- $300 M_\odot$ of stars.  It
has been suggested that mass of the most massive star in a cluster
correlates with the mass of the cluster, with clusters less massive 
than $\sim 250 M_\odot$ being incapable of forming an O-star 
\citep[e.g.][]{Larson82, Weidner06}.  However, \citet{deWit04, deWit05} 
have found that roughly $4$\,\% ($\pm 2$\,\%) of {\em all} O-stars appear 
to have formed in isolation, in that no (significant) host cluster
is present, and they cannot be accounted for as a runaway ejected
star.

The observations of \citet{deWit04, deWit05} can be explained in one of
two ways.  Firstly, it may be that some very low-mass clusters are
able to form O-stars, and therefore there is no limit on
the maximum mass of a star in a low-mass cluster (other than the total
mass of that cluster).  Therefore these isolated O-stars are just the
extreme tail of a distribution of stellar masses in low-mass
clusters \citep[see e.g.][]{deWit05}.  Secondly, if there {\em is} a 
limit on the maximum stellar mass a cluster of a given mass may 
produce, then these isolated O-stars must be formed by a different 
mechanism to the vast majority of stars that form in clusters, or that
the host clusters must rapidly disperse.

In this paper we address the possible origin of these isolated
O-stars.  In Section~\ref{method} we describe our Monte
Carlo methods, and we present our results in Section~\ref{results}. We 
conclude in Section~\ref{conclude}.

\section{Method}
\label{method}

We form a population of clusters and stars by randomly sampling first
from a power-law cluster mass function, and then populating that
cluster from an initial mass function (IMF).

Cluster masses are selected from a power-law cluster mass function
(CMF) of 
the form $N(M) \propto M^{-\beta}$ between a lower and an upper mass 
limit.  The lower mass limit is usually taken to be $M_{c0} = 
50\,M_\odot$, the upper mass limit is allowed to vary, but is usually
in the range $M_{c1} = 10^4$ -- 10$^5\,M_\odot$.  The {\em total} mass 
of clusters in each Monte Carlo run is $10^9\,M_\odot$ in order to
fully sample the mass range of clusters.

\subsection{The stellar IMF}

Each cluster is populated with stars drawn from a 
two-part \citet{Kroupa02} IMF of the form
\begin{displaymath}
 N(M)   \propto 
  \left\{ \begin{array}{ll}
  M^{-1.3} \hspace{0.4cm} m_0 < M/M_\odot < m_1   \,, \\
  M^{-2.3} \hspace{0.4cm} m_1 < M/M_\odot < m_2   \,,
\end{array} \right.
\end{displaymath}
where $m_0$ and $m_2$ are the lower and upper limits of the IMF
respectively, and $m_1$ is the mass at which the IMF
slope changes.  

We choose $m_0 = 0.1\,M_\odot$, and $m_1 = 0.5\,M_\odot$ 
\citep[see][]{Kroupa02}.  We note that whilst brown
dwarfs are numerous, their contribution to the total mass of the
cluster is very small and so we do not include them in our
calculations. 

There are two ways of determining the population of stars
within a cluster.  Firstly, the IMF can be sampled to allow the 
possibility of a small cluster to contain a star close
to the total mass of the cluster.  Secondly, we can limit the maximum
mass of a star in a cluster to be related to the mass of the cluster 
\citep[cf.][]{Weidner04}.

\subsubsection{Random sampling of the IMF}
\label{random}

When randomly\footnote{Formally the cluster is populated in a constrained
  sampling, as the mass of the most massive star cannot exceed the
  mass of the cluster.} sampling the IMF we set $m_2 = 150\,M_\odot$, 
the fundamental upper limit on the mass of stars 
\citep{Oey05,Figer05,Weidner04}.  Recent studies \citep{Oey07} point out that 
this upper mass limit may be as high as $200\,M_\odot$, however we us
the more standard upper stellar mass limit of $m_2 = 150\,M_\odot$ 
throughout this paper.  

Stars are added to a cluster until the {\em total} mass of the stars
in a particular cluster is between 98\,\% and 105\,\% of the mass of
that cluster. If the final star to be added to the cluster exceeds 
the 105\,\% limit then the cluster is {\em entirely} re-populated 
\citep[see e.g.][]{Goodwin05}.

Our sampling technique differs from that used by
  \citet{Elmegreen06} who used a `soft-sampling' method in which the
  final star to be added to the cluster can be of any mass, and thus
  the final mass of the cluster could be greater than the initially 
  sampled mass.  We note that our results differ only negligibly using
  `hard' or `soft' sampling.  This is due to the very low probability
  of selecting an O-star as the last star in an almost fully populated
  cluster.

\subsubsection{Limiting the upper mass limit}

It has been suggested that there is an upper limit to the mass of a
star in a cluster which depends on the mass of that cluster
\citep[e.g.][]{Larson82, Weidner04}.  \citet{Weidner04} parameterise 
the maximum stellar mass within a
cluster, $m_{\rm
  max}$, as a function of the initial (embedded) cluster mass, $M_{\rm
  ecl}$, \citep[see][their Section~2.2]{Weidner04}.  We solve
\citeauthor{Weidner04}'s eqn.~8 numerically to obtain the $m_{\rm
  max} - M_{\rm ecl}$ relationship (as illustrated in figs.~\ref{contour-a}~
\&~\ref{contour-b}).

For a randomly sampled cluster of a given mass, we determine $m_{\rm
  max}$ and set the maximum mass $m_2$ in the IMF
  (eqn.~1) to be $1.1 m_{\rm max}$.  We then proceed as above to populate the 
cluster with stars.  We note that this is not ideal as the actual maximum mass
selected for a given cluster tends to be somewhat smaller than the $m_{\rm
  max}$ determined by the \citet{Weidner04} relationship.
However, as we shall show, the details of this method are unimportant
as limiting the upper mass of a star in a cluster can {\em never} 
reproduce the isolated O-star fraction.

\section{Results}
\label{results}

\citet{deWit04, deWit05} found that $\sim 4$\% of the {\em total}
  number of Galactic O-stars are found in apparent isolation.
\citet{deWit05} found that when random sampling from an IMF that
they are able to reproduce the isolated O-star fraction of $\sim 4$\%
by selecting clusters from a power-law of slope $\beta = 1.7$, and
that selecting from a standard cluster mass function (CMF) with $\beta = 2$
produces too many isolated O-stars.
\citet{deWit05} calculated the fraction of isolated O-stars by
  summing the number of O-stars in clusters that only contain a single
  O-star, and dividing by the total number of O-stars in {\em all} clusters.

Note that throughout the rest of this paper we will use
  the term `isolated O-star', however a better term would be
  `apparently isolated O-star' as, whilst there is not a significant
  population of other stars present around these O-stars, there could be
  (and we suggest there is) a small population of low-mass stars.

Our results are summarised in Table~\ref{table}. Following \citet{deWit05} 
we define an O-star to be a star with mass 
$>17.5\,M_\odot$\footnote{We note that changing the minimum mass of an
  O-star within a reasonable range of values does not change the 
results significantly.}.  We agree with the results of \citet{deWit05}
and find that $\sim 6$\,\% of the total number of O-stars are single when
drawn from a CMF with $\beta = 1.7$, while $\sim 17$\,\% of O-stars are
single for a CMF with $\beta = 2$.

\begin{table*}
\begin{center}
\begin{tabular}{|c|c|c|c|c|}
\hline Stellar Sampling & $\beta$  & Cluster upper mass 
& Maximum number of & Isolated O-star \\ 
 & & limit & B-stars & fraction \\
\hline
random & 1.7 & $M_{c1}$ & Any & 6.0\,\% \\
random & 1.7 & $M_{c1}$ & 0 & 3.1\,\% \\
random & 1.7 & $<100$\,$M_\odot$ & Any & 1.5\,\% \\
random & 1.7 & $<100$\,$M_\odot$ & 0 & 1.3\,\% \\
\hline 
random & 2 & $M_{c1}$ & Any & 16.7\,\% \\
random & 2 & $M_{c1}$ & 0 & 9.7\,\% \\
random & 2 & $<100$\,$M_\odot$ & Any & 5.2\,\% \\
random & 2 & $<100$\,$M_\odot$ & 0 & 4.6\,\% \\
\hline
CMMSM & 2 & $M_{c1}$ & Any & 4.0\,\% \\ 
CMMSM & 2 & $M_{c1}$ & 0 & 0.4\,\% \\ 
CMMSM & 2 & $<100$\,$M_\odot$ & Any & 0\,\% \\
CMMSM & 2 & $<100$\,$M_\odot$ & 0 & 0\,\% \\
\hline
\end{tabular}
\end{center}
\caption[bf]{A summary of the main results. The columns show the
  parameters for (a) the type of stellar IMF 
sampling, (b) the slope of the cluster mass function ($\beta$), (c) the upper 
limit for the mass of a cluster to contain an isolated O-star, (d) the number of 
B-stars allowed in a cluster, and (e) the resulting isolated O-star 
fraction. The first four calculations are for a cluster 
mass function with $\beta = 1.7$ (as adopted by
\citet{deWit05}).  The next four calculations are for a `standard' cluster 
mass function with $\beta = 2$ \citep{Lada03}.  In both cases the
maximum allowable mass of a star in a cluster is the cluster mass
(random sampling).  The final four calculations are for a $\beta = 2$
cluster mass function, but with the maximum mass of a star in a
cluster constrained by the cluster mass-maximum stellar mass (CMMSM) 
relationship.}
\label{table}
\end{table*}

However, many of those single O-stars are present in large clusters
(up to $\sim 10^3 M_\odot$, see fig.~1) which should have been detected 
by \citet{deWit04}.  In addition, many of the clusters containing just a 
single O-star also contain one or more B-stars that would also have
been detected by \citet{deWit04, deWit05} (unless they were close
companions to the O-star).

To account for the lack of other massive (B-) stars and significant
numbers of low-mass companions around isolated O-stars, we 
restrict the definition of an
isolated O-star to be one in which the total cluster mass is $<100
M_\odot$, and one which has no B-stars (defined as stars with masses 
$10 > M/M_\odot < 17.5$).  Hence the fraction of isolated O-stars
  becomes the number of O-stars in small clusters with no B-star
  companions, divided by the total number of O-stars in all clusters.

\begin{figure}
\begin{center}
\rotatebox{270}{\includegraphics[scale=0.33]{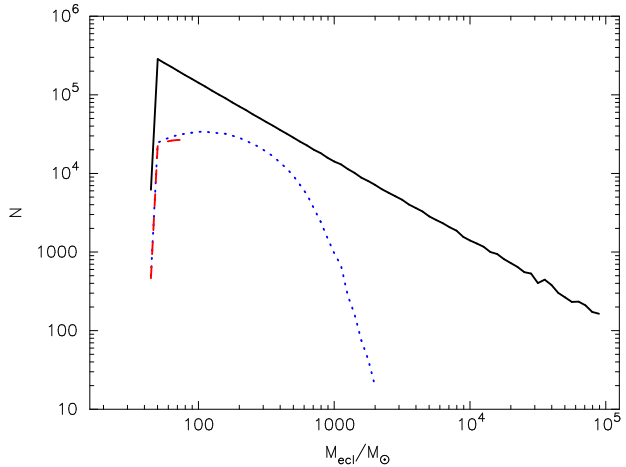}}
\end{center}
\caption[bf]{A plot of the cluster mass functions (CMFs) for a slope of 
$\beta$ = 2. The CMF for all clusters is shown by the black (solid) line; 
the CMF for clusters containing only one O-star is shown by the blue (dotted) 
line; and the CMF for clusters containing one O-star, no B-stars and a cluster 
mass $M_{\rm ecl} <$ 100\,M$_\odot$ is shown by the red (dashed) line.}
\label{cmfs}
\end{figure}

The effect of including these extra restrictions is dramatic.  Fig.~\ref{cmfs}
shows the $\beta = 2$ CMF of all clusters (solid line), as well as the CMF of
clusters containing a single O-star (dotted line), and the CMF of
clusters $<100M_\odot$ containing a single O-star and no B-stars (dashed line).  
The fraction of O-stars that are `isolated' falls from $16.7$\,\%
to $4.6$\,\%.  Note that this agrees with the observation by \citet{Oey04} 
showing that (in the SMC) a power-law of $\beta = 2$
describes the cluster richness distribution down to clusters
containing single O-stars.  For $\beta = 1.7$, however, the fraction
of isolated O-stars falls to $1.3$\,\% (with the same restrictions).

The results are not very sensitive to the maximum mass of a cluster
within which an O-star is considered isolated (taken to be $100
M_\odot$).  Cluster masses much in excess of $100 M_\odot$ often
also include B-stars and so are discounted on those grounds, and
clusters less massive than $\sim 50 M_\odot$ almost never produce an
O-star (indeed, only $\sim 10$\,\% of $100 M_\odot$ clusters
ever produce an O-star).

The fraction of isolated O-stars is slightly sensitive to
the upper mass limit of the CMF.  When the upper mass limit is changed
from $10^4$ to $10^5 M_\odot$ the fraction of isolated O-stars changes
from $6.8$\,\% and $4.6$\,\% for $\beta = 2$ and between $3.0$\,\% and
$1.3$\,\% for $\beta = 1.7$ (due to the increasing number of very
massive clusters that are able to fully sample the IMF up to the $150
M_\odot$ stellar mass limit).

Thus a standard CMF with $\beta = 2$ as observed for clusters in the
Solar Neighbourhood \citep{Lada03} is able to reproduce the
isolated O-star fraction when reasonable limits are placed on the
presence of other massive (i.e. B-) stars, and on the total stellar
mass that can be associated with the O-star (usually less than $\sim
80 M_\odot$ of M-, K-, and G-dwarfs).

This suggests that the
only limit on the most massive star in a cluster is whichever is
smaller of the cluster mass and the fundamental upper-limit of stellar
masses (which we take to be $150 M_\odot$).

The random sampling model predicts that there should be a population 
of single O-stars in fairly small clusters of a few hundred $M_\odot$.  
\citet{deWit04, deWit05} find a similar fraction of their field O-star
sample lie in small clusters as are found to be isolated (see tables~1
and~3 in \citealt{deWit05}, $12/43$ associated with clusters compared
to $11/43$ isolated).  The fraction of single O-stars with no B-stars
in clusters of mass $100< M_{\rm ecl} < 300-500\,M_\odot$ drawn from a
CMF with $\beta = 2$, is $\sim$ 4\%, 
consistent with the fraction of single O-stars in modest clusters
found by \citet{deWit04,deWit05}.

\subsection{The cluster mass-maximum stellar mass relation}
\label{CMMSM}

It has been suggested that there is a cluster mass-maximum stellar
mass (CMMSM) relationship (\citealt{Larson82,Weidner06}; however see 
\citealt{Elmegreen05}). In order to contain a 
star of $>17.5\,M_\odot$ a cluster must be more massive 
than $\sim 300\,M_\odot$ \citep{Weidner06}.

Clearly, in such a situation, {\em no} O-stars could fulfil our
criteria of `isolation'.  Monte Carlo simulations of a CMF with $\beta
= 2$ and limiting the maximum mass of a star within a cluster
according to the \citet{Weidner06} relationship (see Section~\ref{random})
gives a single O-star fraction of $\sim 4$\,\%.  However, including
our extra constraints that no B-stars are present reduces this
fraction to $0.4$\,\%, and further adding the constraint that the
maximum mass of the cluster is $<100 M_\odot$ (unsurprisingly) reduces
this fraction to zero.

\subsection{Is there a cluster mass-maximum stellar mass relationship?}
\label{DCMMSM}

Thus, if there exists a CMMSM relationship,
the population of isolated O-stars apparently cannot be accounted for
within the clustered mode of star formation.  We are left with two  
possibilities.  Firstly, that isolated O-stars form from a different
mode of star formation to the dominant clustered mode.  Secondly, that
stars do form in clusters with the CMMSM relationship, but some of
those clusters rapidly disperse leaving an isolated O-star \citep[e.g. due 
to low star formation efficiencies and then gas expulsion, see][]
{Bastian06,Goodwin06}.  We note however, that the distribution of star 
formation efficiencies would have to be  such as to mimic the results of 
random sampling from the CMF which would presumably require fine-tuning.
 
\begin{figure}
  \begin{center}
\rotatebox{270}{\includegraphics[scale=0.33]{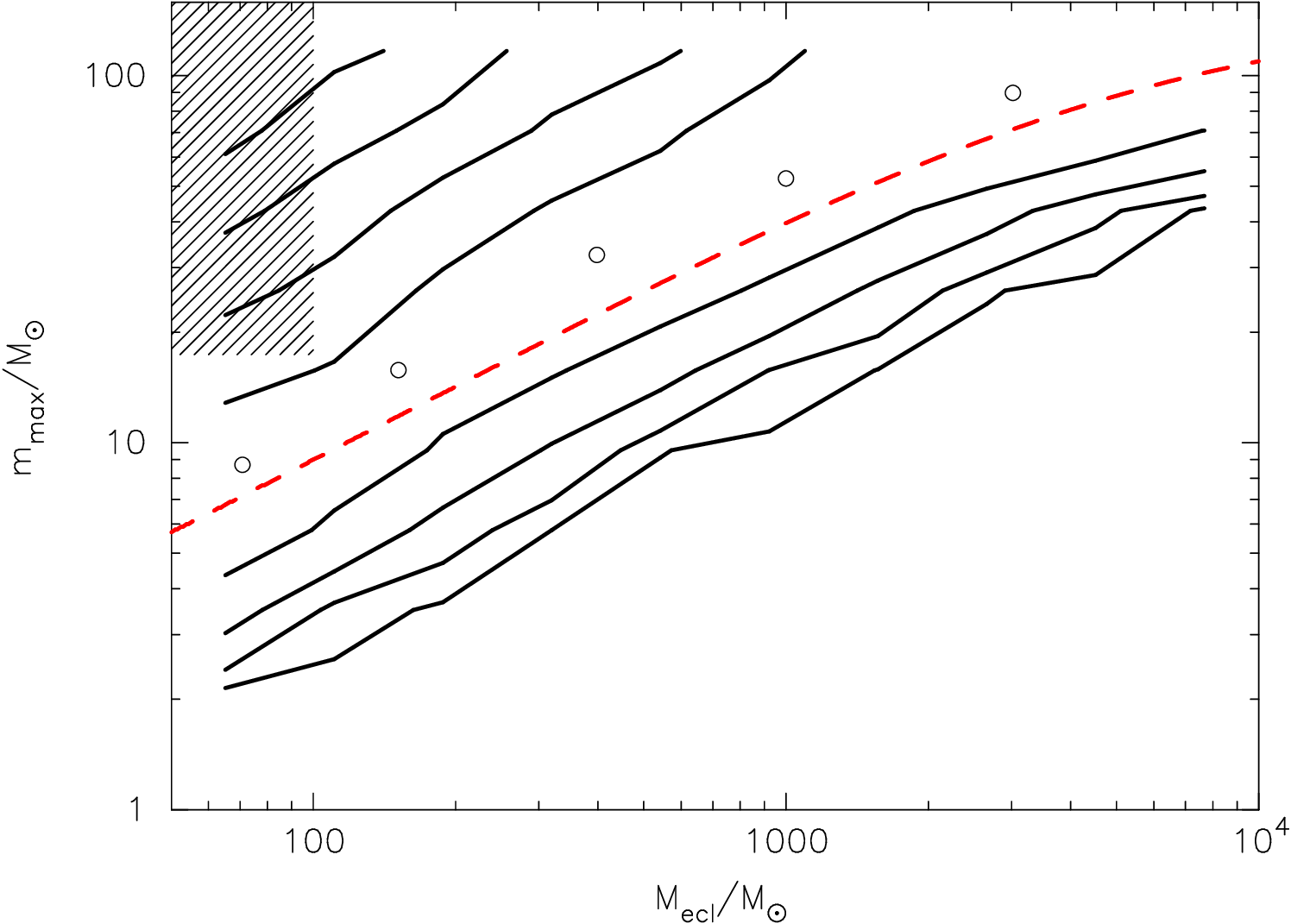}}
  \end{center}
  \caption[bf]{A contour plot of the numbers of clusters of 
  mass $M_{\rm ecl}$ having a most massive star of mass $m_{\rm
  max}$ drawn from a uniform ($\beta = 1$) cluster mass function.  
\citeauthor{Weidner06}'s \citeyearpar{Weidner06} relationship 
between $M_{\rm
  ecl}$ and $m_{\rm max}$ is shown by the red dashed line.  The region
  in which isolated O-stars would be found (i.e. $m_{\rm
  max} > 17.5 M_\odot$ and $M_{\rm ecl} < 100 M_\odot$) is indicated
  by the hashed region in the top left of the figure.  The contour levels from 
the CMMSM line outwards are 200, 100, 40 and 10 clusters (from a total of 
9248 clusters). The median stellar mass for clusters in a mass bin range of 
0.2 dex is also shown by the open circles.}
  \label{contour-a}
\end{figure}

In fig.~\ref{contour-a} we show the numbers of clusters of a
given mass ($M_{\rm ecl}$) harbouring a star of a given maximum mass
($m_{\rm max}$).  In order to remove the potentially confusing 
  effect of the CMF, in fig.~\ref{contour-a} we plot the contours for a uniform
  ($\beta=1$) CMF.  The hashed region corresponds approximately to 
our definition of an isolated O-star\footnote{The region shows 
$m_{\rm max} > 17.5 M_\odot$ and $M_{\rm ecl} < 100 M_\odot$, but
  we are unable to also illustrate the constraint that there are no
  B-stars.  However, the no B-star constraint has little effect in
  low-mass clusters.}.  The dashed-line shows the CMMSM relationship 
from \citet{Weidner06}.  Interestingly, the CMMSM relationship falls
a little below the {\em median} of the distribution (shown by the open 
circles), suggesting that the CMMSM relationship may simply be the 
`typical' result of star formation in clusters.

In order to test the possibility that the CMMSM relationship reflects
`average' clusters, we pre-select a set of $16$ clusters with
masses very similar to those in the observed sample of \citet[see e.g. 
their fig. 7/table 1]{Weidner06}.  We then populate these
clusters randomly from the IMF and examine how often the most massive
stars in each of the clusters lie close to the CMMSM relationship.  We
find that only $\sim 10$\,\% of the time all of the 16 clusters have a
CMMSM relationship that is within $0.25$ dex of the `true' CMMSM
relationship.  In $90$\,\% of trials at least one (most often one or 
two) cluster lies more than $0.25$ dex from the relationship.

This might suggest that the CMMSM relationship represents more than
the `average' cluster.  However, in almost all cases
the clusters that lie far from the CMMSM relationship are low-mass
clusters that contain an overly massive star.  That is, clusters that
do not fit the CMMSM relationship lie in or around the `isolated
O-star' region of fig.~\ref{contour-a}.

\begin{figure}
  \begin{center}
\rotatebox{270}{\includegraphics[scale=0.33]{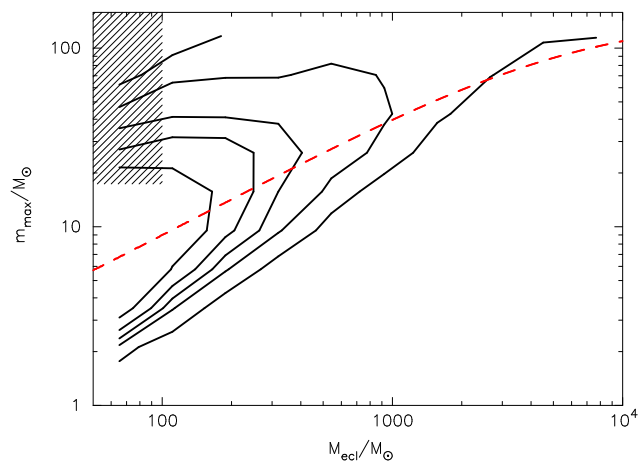}}
  \end{center}
  \caption[bf]{As fig.~\ref{contour-a} but for cluster masses drawn
  from a $\beta~=~2$ power-law cluster mass function. The contour levels from 
the CMMSM line outwards are 400, 250, 150, 70 and 30 clusters (from a total 
of 8513 clusters).} 
  \label{contour-b}
\end{figure}

In fig.~\ref{contour-b} we show the numbers of clusters with a mass
$M_{\rm ecl}$ and a maximum stellar mass $m_{\rm max}$ when clusters
are drawn from a power-law with slope $\beta = 2$.  Low-mass clusters
are by far the most common clusters, and the most common deviations
from the CMMSM relationship are towards the top left of the diagram -
in the region populated by isolated O-stars.  Therefore, the clusters 
most likely to fall significantly away from the
CMMSM relationship appear to be isolated O-stars and are therefore not
included in any CMMSM relationship.

\subsection{Other massive stars}
\label{other_stars}

In a series of papers, 
\citet{Testi97} and \citet*{Testi98,Testi99} analysed infra-red 
observations of clusters surrounding Herbig Ae/Be stars. They define a 
`richness indicator' -  proportional to the number of stars in the surrounding 
cluster and plot this as a function of spectral type 
\citep[see][their fig.~3]{Testi99}.  From inspection of fig.~3 
in \citet{Testi99}, the fraction of B-stars in very modest clusters
appears to be around 5\% (although with very large error bars).

We can define an `isolated B-star' in a similar way 
to isolated O-stars: a single star of mass $10\,M_\odot < M_{\rm B\star} \le
17.5\,M_\odot$, within a cluster of mass $M_{\rm ecl} < 100 M_\odot$ 
which also contains no O-stars.  With such a definition we can predict
that there should be an isolated B-star fraction of $\sim
6$\,\%, somewhat higher than (but within the error bars of) the 
isolated O-star fraction, and consistent with \citet{Testi99}.   

\begin{figure}
\begin{center}
\rotatebox{270}{\includegraphics[scale=0.33]{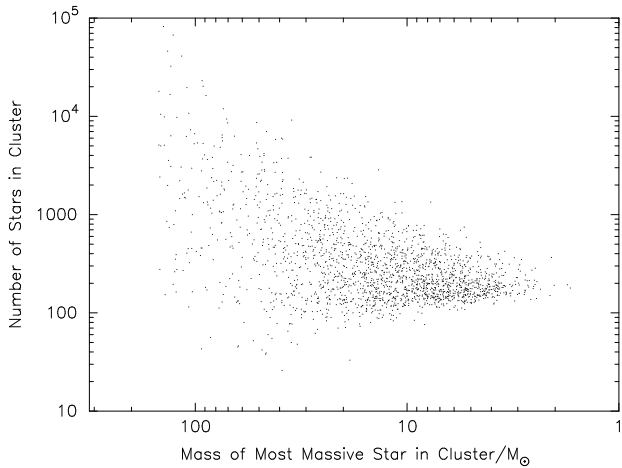}}
\end{center}
\caption[bf]{The number of stars in a cluster plotted as a function of the 
most massive star in that cluster for a random sampling of the IMF. }
\label{testi_random}
\end{figure}

\begin{figure}
\begin{center}
\rotatebox{270}{\includegraphics[scale=0.33]{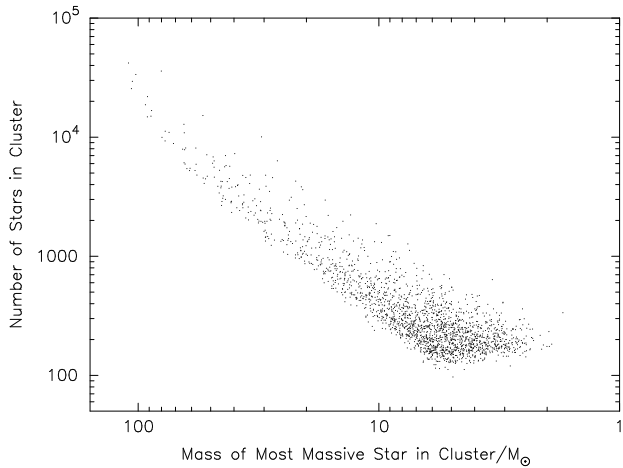}}
\end{center}
\caption[bf]{The number of stars in a cluster plotted as a function of the 
most massive star in that cluster for a fixed sampling of the IMF using the 
CMMSM relation advocated by \citet{Weidner04,Weidner06}. }
\label{testi_wk}
\end{figure}

We model the analysis of \citet{Testi99} by plotting the number of stars in 
a particular cluster as a function of the most massive star in that cluster 
for two scenarios. Firstly, we populated clusters randomly from the
IMF, and secondly using the CMMSM relation.  Figs.~\ref{testi_random}~and~
\ref{testi_wk} show the
distribution of cluster richness (the total number of stars in each
cluster) for the random and CMMSM sampling respectively.  Comparison
of figs.~\ref{testi_random}~and~\ref{testi_wk} with fig.~3 from 
\citet{Testi99} shows that the random sampling reproduces the observations 
far better.  We note again that the rapid dispersal of some clusters would 
add a scatter to the initially strong CMMSM relationship, however, once more 
the dispersal of clusters would have be fine-tuned to match the
predictions of the random sampling model.

\section{Conclusions}
\label{conclude}

We use Monte Carlo simulations to form clusters from a power-law
cluster mass function (CMF) and then populate these clusters with
stars from a stellar initial mass function (IMF).  We sample randomly
from a \citet{Kroupa02} IMF with an upper limit of $150 M_\odot$ or with
an upper limit set from the cluster mass-maximum stellar mass (CMMSM)
relationship from \citet{Weidner06}.

\citet{deWit04, deWit05} find that $4 \pm 2$\,\% of O-stars ($M >
17.5 M_\odot$) form in apparent isolation, i.e. with no associated
cluster, no other massive stars, and with no apparent ejection from a
nearby cluster.

We agree with \citet{deWit05} that from a CMF with slope $\beta =
1.7$ around $4$\,\% of O-stars form in clusters that contain no other
O-stars, and that for a `standard' CMF of $\beta = 2$ \citep[e.g.][]{Lada03} 
this fraction is $\sim 17$\,\%.  

When we include additional constraints that an isolated O-star must also have no
B-stars in the same cluster, and must form in a cluster of $< 100
M_\odot$ (in order to match the observational constraints from 
\citet{deWit04, deWit05}), we find that the fraction of isolated O-stars for 
a CMF with $\beta = 2$ falls to $\sim 5$\,\% in-line with the
observations, whilst for $\beta = 1.7$ the fraction falls to $\sim
1$\,\%.  This suggests that the isolated O-stars can be
explained from a standard CMF if stellar masses are drawn randomly
from an IMF, and therefore very low-mass clusters do have a (small)
probability of forming a very massive star.

If there is a cluster mass-maximum stellar mass (CMMSM) relationship
\citep[e.g.][]{Larson82, Weidner06} then there is {\em no}
chance of isolated O-stars forming in clusters as the
lowest mass cluster that can form an O-star is $\sim 275 M_\odot$.
The CMMSM relation also appears not to hold for clusters in which the most 
massive star is of spectral type B or A. We find that we can reproduce 
the observations by \citet{Testi99} if we randomly sample the IMF, whereas 
introducing the CMMSM relation into our simulations produces something 
altogether different. 

If the CMMSM relationship is physical, then either (a) isolated
O-stars must form from a different mechanism to the bulk of stars that
form in clusters with a power-law slope of $\beta \sim 2$, or (b)
clusters containing single O-stars must disperse rapidly leaving
isolated O-stars.  Option (a) is unsatisfying as it requires a
different mode of star formation for a small fraction of the most
massive stars.  Option (b) is difficult to reconcile with the
observations.  The most probable cause for the rapid dispersal of some
clusters is that they form with a low star formation efficiency and
are destroyed after gas expulsion \citep[see e.g.][]{Bastian06,Goodwin06}.  
However, the star formation efficiencies of
clusters must be fine-tuned so that they match the predictions of
purely randomly sampling from a standard CMF.  In addition, single
O-stars form in clusters with large numbers of B-stars, and so the
fields around apparently isolated O-stars should also contain many
B-stars.  Whilst this has not been actively searched for, there is no
obvious sign of many isolated B-stars in the survey of \citet{deWit04}.

Therefore we suggest that the CMMSM relationship is
an average relationship between the most common maximum stellar
mass within a cluster of a given mass.  In a sample of clusters
chosen to mimick the sample of \citet{Weidner06}, the CMMSM
relationship is only recovered in $\sim 10$\,\% of random populations
of those clusters.  However, most extreme deviations from the CMMSM
relationship are very low-mass clusters which form a particularly
massive stars: i.e. an `isolated O-star', which are excluded from 
the CMMSM relationship as their 
identification with very low-mass clusters has previously been 
unclear.

We note that young stars, of all masses, 
tend to be X-ray bright \citep[see e.g.][]{Feigelson99, Ramirez04, Getman06}.  
Therefore X-ray observations of the regions around isolated O-stars could 
provide an indication of the presence of associated young low-mass
stars around apparently isolated O-stars.

In summary, the existence and number of isolated 
O-stars \citep{deWit04, deWit05} can be explained within a standard 
cluster mass function with $\beta = 2$ as long as very
low-mass clusters are able to form massive stars 
\citep[e.g.][]{Elmegreen05,Elmegreen06}.  

\section*{Acknowledgements}

RJP acknowledges financial support from PPARC/STFC.  SPG 
acknowledges the support and hospitality of the International Space
Science Institute in Bern, Switzerland where part of this work was
done as part of a International Team Programme. We thank the anonymous 
referee for their comments on the original manuscript. Thanks also to Francis Keenan, 
Danny Lennon and Chris Evans for information on the fraction of isolated B-stars; 
and to Pavel Kroupa and Thomas Maschberger for useful discussions and suggestions.

\bibliographystyle{mn2e}
\bibliography{reference}

\label{lastpage}

\end{document}